\documentclass[preprint,amssymb,amsmath,aip,cha]{revtex4-1}
\usepackage{graphicx}
\usepackage[caption = false]{subfig}
\usepackage{color}
\usepackage{epsfig}
\usepackage{ifpdf}
\usepackage{bm}
\usepackage[colorlinks=true,linkcolor=blue]{hyperref}%
\expandafter\ifx\csname package@font\endcsname\relax\else
 \expandafter\expandafter
 \expandafter\usepackage
 \expandafter\expandafter
 \expandafter{\csname package@font\endcsname}%
\fi
\usepackage{cleveref}
\begin{document}
\title{Influence of External Magnetic Field on Dust$-$Acoustic Waves in a Capacitive RF Discharge}
\author{Mangilal Choudhary}
\email{Mangilal.Choudhary@exp1.physik.uni-giessen.de} 
\affiliation{I. Physikalisches Institut, Justus--Liebig Universität Giessen, Henrich--Buff--Ring 16, D 35392 Giessen, Germany}
\author{Roman Bergert}
\author{Slobodan Mitic}
\author{Markus H. Thoma}
\begin{abstract}
This paper reports experiments on self$-$excited dust acoustic waves (DAWs) and its propagation characteristics in a magnetized rf discharge plasma. The DAWs are spontaneously excited in dusty plasma after adding more particles in the confining potential well and found to propagate in the direction of streaming ions. The spontaneous excitation of such low-frequency modes is possible due to the instabilities associated with streaming ions through the dust grain medium. The background E-field and neutral pressure determine the stability of excited DAWs. The characteristics of DAWs strongly depend on the strength of external magnetic field. The magnetic field of strength B $<$ 0.05 T only modifies the characteristics of propagating waves in dusty plasma at moderate power and pressure, P = 3.5 W and p = 27 Pa respectively. It is found that DAWs start to be damped with increasing the magnetic field beyond B $>$ 0.05 T and get completely damped at higher magnetic field B $\sim$ 0.13 T. After lowering the power and pressure to 3 W and 23 Pa respectively, the excited DAWs in the absence of B are slightly unstable. In this case, the magnetic field only stabilizes and modifies the propagation characteristics of DAWs while the strength of B is increased up to 0.1 T or even higher. The modification of the sheath electric field where particles are confined in the presence of the external magnetic field is the main cause of the modification and damping of the DAWs in a magnetized rf discharge plasma.
\end{abstract}
\maketitle
\section{Introduction}
The presence of submicron to micron-sized particles in a plasma makes it more complex because these particles alter the dynamics of the plasma species (electrons and ions) as well as they exhibit their own dynamics. Such medium, which consists of three charged species namely electrons, ions, and charged solid particles, is termed as a dusty plasma or complex plasma. In the background of a low-temperature plasma, energetic electrons impinge on the surface of the solid particle and charge their surface negatively up to $10^3-10^5$ times of an electron charge\cite{Charging} to balance the fluxes of electrons and ions. After the density of negatively charged dust particles crosses a critical value, then long-range coloumbic interaction among the dust particles turns the dust grain medium to respond collectively similar to the plasma. The results of the collective phenomena of dusty plasma are predicted theoretically and experimentally in the form of linear and non-linear waves such as longitudinal dust acoustic waves \cite{raodaw1,daw2,daw3,mangilalpop} dust acoustic transverse waves \cite{lmode,tsw,vikramtsw}, dust lattice waves \cite{dlw1,dlw2,tdlw}, dust acoustic solitary waves \cite{kdv,pdasw} and dust acoustic shock waves \citep{dasw,expdasw,exp1dasw} as well as dust vortices\cite{mangilargeaspect,modhuvortices,mangilalmultiplerot} \\
The study of dust acoustic waves (DAWs) has a more than 25 years old history \cite{dawmerlino}. In low-temperature discharges (DC or RF), the DAWs are excited in the dust grain medium which is confined in a 
strong electric field (E) of the cathode sheath\cite{mangilalpop,pdasw} or anodic plasma\cite{daw2} or rf sheath \cite{rfdischarge,cristopherbcc,benirfdischarge} or diffused plasma \cite{icpddw, mangilalmultiplerot,mangilargeaspect}. Such low frequency dust acoustic modes are either excited by inherent instabilities such as ion-streaming instability and dust--dust instability \cite{instability1,instability2,instability3,instability5} or externally forced triggered instabilities\cite{Schwabeexternalwxcired, bruamultisoliton}. However, the ion-streaming instability which arises due to the streaming of ions relative to the stationary dust particles is considered as a main free energy source to excite DAWs in DC or RF discharge. The amount of free energy depends on the velocity of the streaming ions ($v_i = \mu_i E$) through the dust grain medium, where $\mu_i$ is the mobility of ion. In the laboratory, both of the variables ($E$ and $\mu_i$) can be externally controlled by changing the plasma or discharge conditions. Previous experimental studies on the excitation and damping of DAWs were performed in the unmagnetized DC or RF plasma background \cite{dawmerlino}. Theoretical studies on DAWs for a magnetized plasma background have been reported in refs.\cite{dawmagnetized1,dawmagnetized2}. In the plasma, an external magnetic field (B) is considered as a variable which modifies the dynamics of electrons and ions. Since the dust particles dynamics is associated with the plasma background, the propagation characteristics of DAWs is expected to be modified in the magnetized plasma. There is a lack of experimental work reported on the dust acoustic waves in a magnetized rf discharge. Therefore, the influence of the external magnetic field on the longitudinal DAWs in a capacitively coupled discharge has been the subject matter of present study.\par

The propagation characteristics of DAWs in the presence of an external magnetic field in a capacitively coupled discharge plasma are discussed. The dust grains are confined in the sheath region of the lower electrode and form a stable dusty plasma. The dusty plasma becomes unstable above a threshold dust number density at low background gas pressure (p $<$ 28 Pa). The inherent instabilities trigger the low-frequency acoustics modes, which propagate along the direction of streaming ions and gravity. It is found that the magnetic field quenches the dust--acoustic waves, which are excited at moderate rf power (P = 3.5 W) and pressure (p = 27 Pa). However, waves excited at low pressure (23 Pa) and power (P = 3 W) remain with significant modifications in their characteristics in the presence of an external magnetic field. The quenching and modified characteristics of DAWs are explained on the basis of the modified sheath electric field and dust charge in the presence of an external magnetic field.
The manuscript is organized as follows: Sec.~\ref{sec:exp_setup} deals with the detailed description of the experimental setup and diagnostics for the dusty plasma studies. Sec.~\ref{sec:DAW without B} describes the excitation and characteristics of DAWs without external magnetic field. The propagation characteristics of DAWs in the presence of the external magnetic field is discussed in Sec.~\ref{sec:daw with B}. The discussion on the observed results is provided in Sec.~\ref{sec:discussion}. A brief summary of the work is provided in Sec.~\ref{sec:summary}.

\section{Experimental Setup} \label{sec:exp_setup}
\begin{figure*}
 \centering
\subfloat{\includegraphics[scale=0.450]{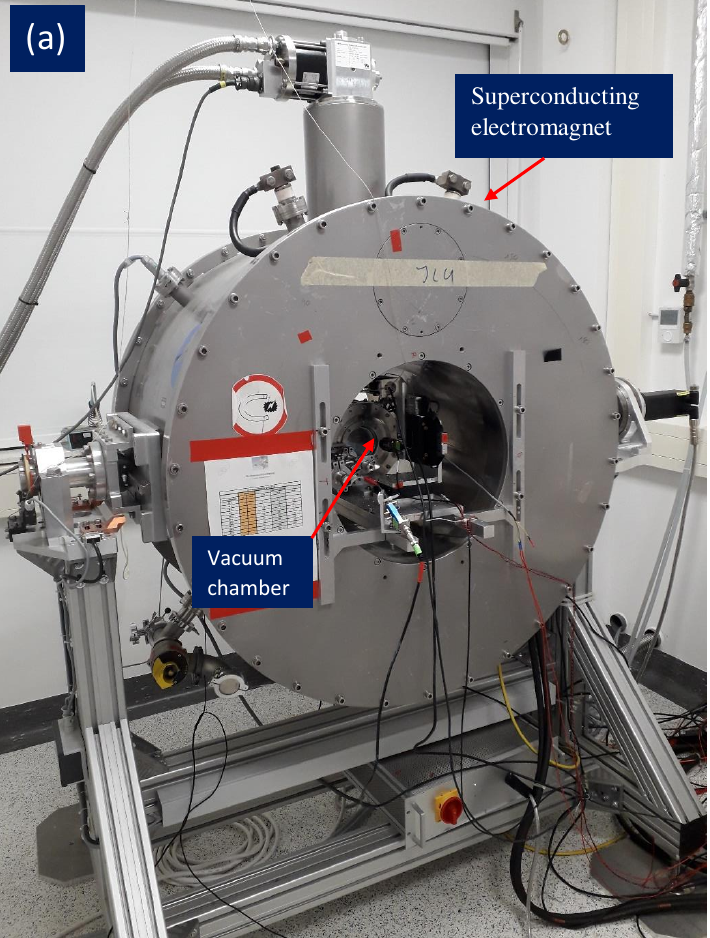} }
\qquad
\subfloat{\includegraphics[scale=0.50]{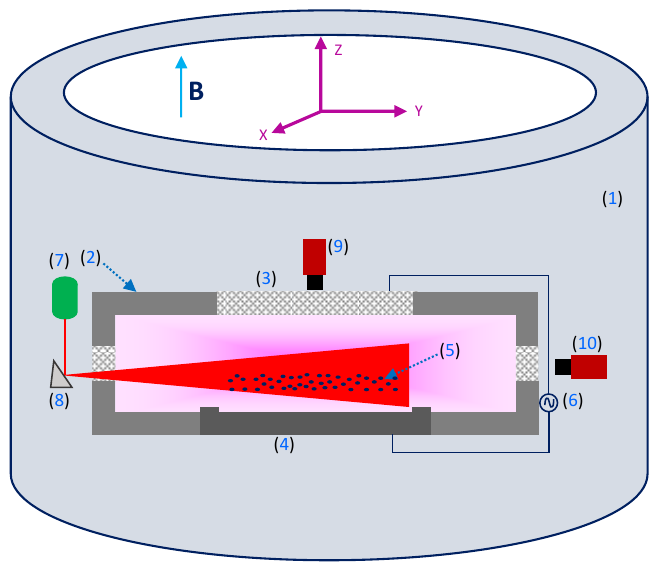} }
 \caption{\label{fig:fig1} (a) Magnetized dusty plasma device. (b) Schematic diagram of the experimental setup. (1) Full view of super conducting magnet, (2) Vacuum chamber, (3) ITO Coated (upper) electrode (4) aluminium (lower) electrode (5) levitated dust particles (6) 13.56 MHz RF power source, (8) Laser, (8) mirror, (9) CMOS camera for horizontal view, and (10) CCD camera for the vertical view.}
 \end{figure*}

 Experiments are performed in an aluminium vacuum chamber, which is placed in the center of a superconducting electromagnet ($B_{max}$ $\sim$ 4 T) to introduce the homogeneous magnetic field in the plasma volume. The magnetized dusty plasma device which is used in the present study is shown in Fig.\ref{fig:fig1}(a). The schematic diagram of the experimental setup is presented in Fig.\ref{fig:fig1} (b). More details about the superconducting electromagnet can be found in ref.\cite{mangilalpsst}. Before the experiments, vacuum chamber is evacuated to base pressure p $<$ $10^{-2}$ Pa using a pumping system consisting of a rotary and turbo molecular pump. A gas mass flow controller (MFC) and gate valve controller are used to control the argon gas pressure inside the vacuum chamber during the experiments. For a given argon pressure, plasma is ignited between an aluminium electrode of 65 mm diameter (lower electrode) and an indium tin oxide (ITO) coated electrode of 65 mm diameter (upper electrode) using a 13.56 MHz rf generator with a matching network. Both electrodes are separated by 30 mm and upper electrode is grounded along with the vacuum chamber. Special design of lower electrode \cite{cristopherbcc}, having a ring shaped periphery of height 2 mm and width of 5 mm, provides a radial confinement to the negatively charged dust particles which are levitated over the lower electrode. The vacuum chamber has 8 side (or radial) ports with total volume of $\sim$ 1000 $cm^3$ . Two opposite side ports are used to diagnose the dusty plasma using a laser and CCD camera in a vertical plane. A CMOS camera is also installed to observe the dust dynamics in the horizontal plane through the transparent upper electrode. A dust dispenser, which is installed at one of the side ports of vacuum chamber, is used for injecting the dust particles into the plasma.
\section{Dust--acoustic waves and its characteristics in absence of magnetic field} \label{sec:DAW without B}
\begin{figure*}
 \centering
\subfloat{\includegraphics[scale=0.850]{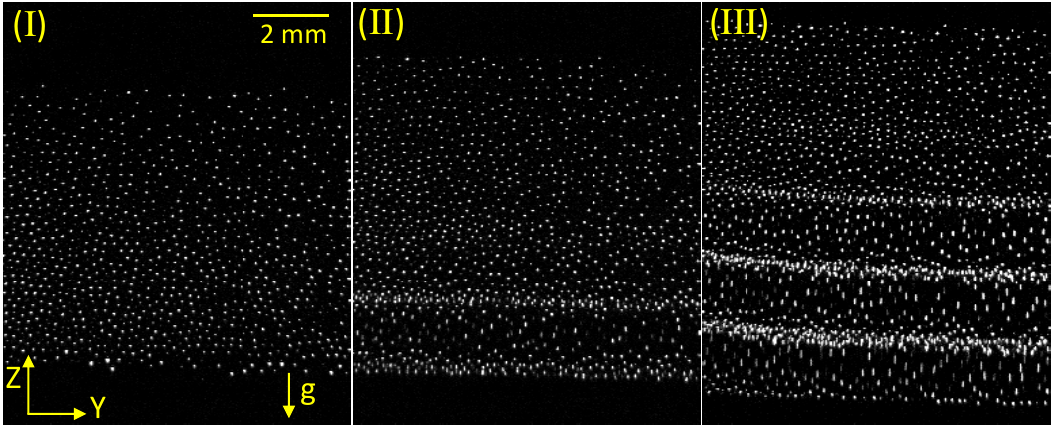} }
\qquad
\vspace*{-0.23in}
\subfloat{\includegraphics[scale=0.450]{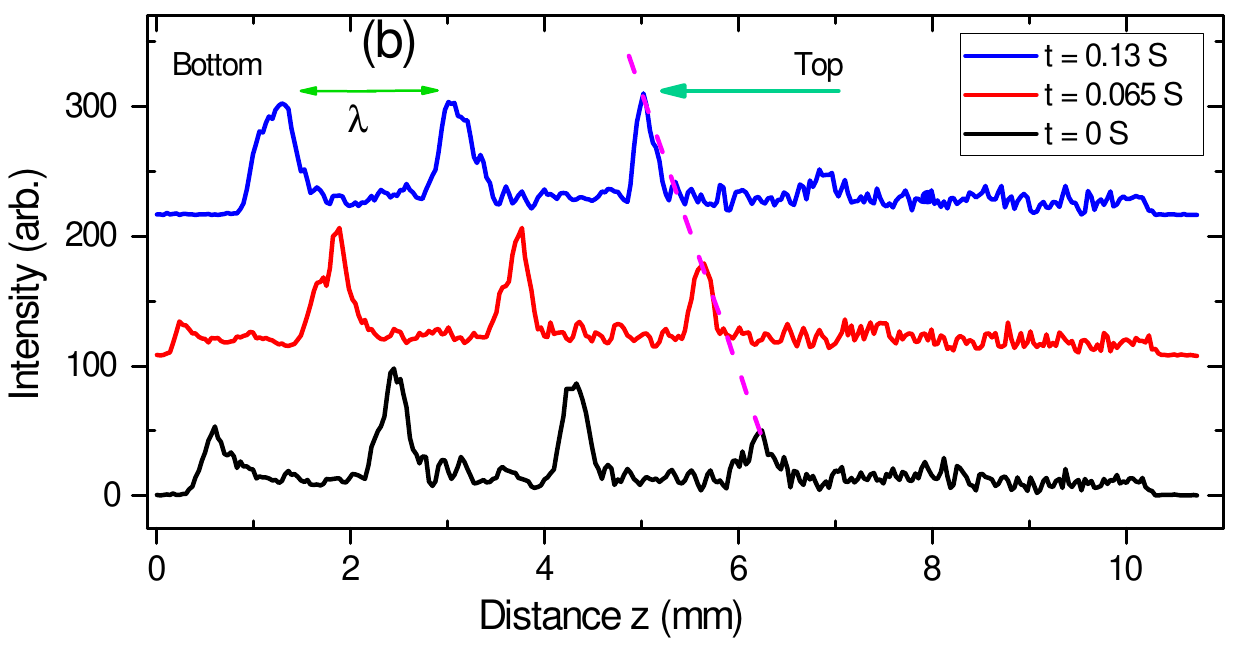} }
\qquad
\subfloat{\includegraphics[scale=0.400]{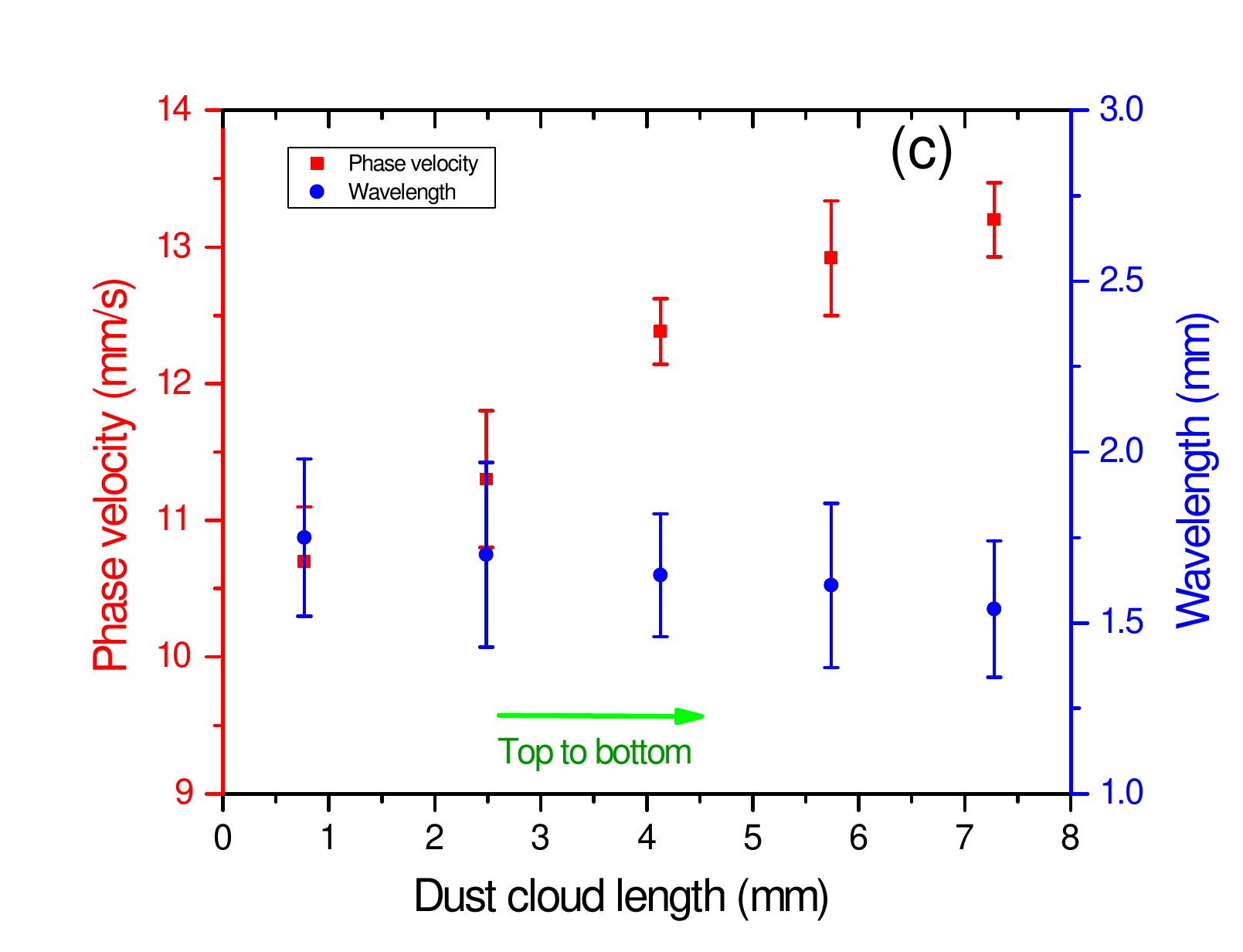} }
 \caption{\label{fig:fig2} Dust grain dynamics in a vertical (Y--Z) plane at center (X = 0) with different dust grain volume. (b) Representation of the time evolution of intensity profile of video frames taken at time step of 65 ms in the Y--Z plane. (c) Wavelength and phase velocity of DAWs from top to bottom in the dusty plasma medium at B = 0 T. The argon pressure and rf power are set to 27 Pa and 3.5 W respectively during the experiments}.
 \end{figure*}

For creating a dusty plasma in the laboratory, mono-dispersive Melamine Formaldehyde (MF) particles of radius, $r$ = 1.025 $\mu$m are introduced using dust dispenser in a capacitively coupled (RF) argon plasma. They get negatively charged in the plasma and confined at the sheath edge of the lower electrode where the dust particles achieve an equilibrium position under the action of various forces \cite{nitterlevitation,iondragforce}. To record the dynamics of the confined dust particles, a red laser to illuminate the dust and a CCD camera to record the scattering light from the grains are used. The CCD camera used in the present sets of the experiment is able to record 20 fps at the resolution of 1024 $\times$ 768 pixels. These stored images are further analyzed using Matlab based code and ImageJ software \cite{imagejsoftware}. It should be noted that the excitation of waves in the dusty plasma is only possible when the dust cloud width exceeds the wavelength of dust--acoustic wave at small dust--neutral friction or low gas pressure. In the laboratory, this is achieved at low-pressure large volume dusty plasma. It is well known that the gravitational force acting on massive particles mainly determines the volume of the dusty plasma. At ground-based experiments, a large volume dusty plasma is possible either in specific discharge configuration \cite{mangirsiexpsystem,mangilargeaspect} or discharges with nano to sub-micron sized particles\cite{nanodusty}. The computer simulation by Chutov and Goedheer \cite{dustyrfplasma} suggests an rf sheath expansion in the presence of a large number of particles. To keep this idea in mind, we have used particles with size $r$ = 1.025 $\mu$m to create a volumetric dusty plasma at low pressure. Since the dust grains exhibit a rotational motion near the outer edge of the lower electrode, the central region of dusty plasma is considered for the study of wave phenomena. For such a study, a 2D vertical plane (Y--Z plane) at the centre (X = 0 cm) is chosen to record the dynamics of particles at various discharge conditions. In this Y--Z plane, dusty plasma is observed to be stable in the case of insufficient free energy to excite the instabilities. It is the case when fewer particles are confined in the potential well, as shown in Fig.~\ref{fig:fig2}(I). After adding more particles into the confining potential well, the number of layers (or dusty plasma volume), as well as dust density increases and the medium, starts to turn into wave-like motion, which is shown in Fig.~\ref{fig:fig2}(II) and  Fig.~\ref{fig:fig2}(III). The instabilities first appear in the lower part (or bottom) of dust grain medium and after that in the upper part (or top). The instabilities arise after adding more particles are a possible source of free energy for the excitation of waves. The growth of instabilities depends on the $E/p$ ratio\cite{instability1}, which increases with the increase of electric field ($E$) or decrease of gas pressure ($p$). Actually, this ratio determines the velocity of streaming ions through the confined dust grain medium, which is considered as the main source of free energy \cite{instability1,instability2,instability5,instability3}.\par

These self-excited waves propagate in the direction of streaming ions at given discharge conditions. The stability of the self-excited DAWs in discharges depends on the $E/p$ ratio, which determines the streaming ion flux through the confined dust grain medium \cite{instability1,instability2}. At low pressure p $<$ 20 Pa, the energy gain of ions is quite higher than dissipation arising from neutral pressure and hence, excited waves are highly unstable. At high pressure p $>$ 30 Pa, the energy loss of ions with neutrals is higher than the energy gain in the sheath electric field and DAWs can not be excited. Moreover, the DAWs stability also depends on E-field, which is connected to input rf power. At higher power P $>$ 4 W, dust grains are confined in a strong E-field region with less number of layers (low volume) and DAWs are observed to be highly unstable. In view of this, we have chosen a pressure and power range of 23 Pa to 28 Pa and 3.5 W to 3 W, respectively for the present study of DAWs. For the detailed characterization of DAWs, the time evolution of the average intensity (or $n_d$) of consecutive still images are considered. Fig.~\ref{fig:fig2}(b) represents the time evolution of average intensity profile of DAWs in the absence of B.  Maxima of the intensity corresponds to the wave crest and minima represents the wave trough. Distance between two consecutive maxima gives the wavelength of DAW. The pink dashed line in Fig.~\ref{fig:fig2}(b) indicates the trajectory of a particular wave crest and this estimates the phase velocity of DAWs. The characteristics of propagating DAWs at argon pressure, p = 27 Pa and rf power, P = 3.5 W are depicted in Fig.~\ref{fig:fig2}(c). It is found that the wavelength ($\lambda$) of DAW decreases from top to bottom and the velocity ($C_{da}$) increases from top to bottom. The frequency of the waves ($f_d = C_{da}/\lambda$) remains almost unchanged in the unmagnetized case (B = 0 T). The variation of particles density ($n_d$) as well as of the E-field throughout the dust grain medium (from top to bottom) causes the variation in the wavelength, $\lambda\sim$ 1.9 mm to 1.3 mm and phase velocity, $C_{da} \sim$ 10.5 mm/s to 13.5 mm/s of DAWs, while they propagate towards the lower electrode. The frequency of DAWs comes out to be $\sim$ 7 Hz. The average phase velocity of DAW is $\sim$ 12 mm at the mentioned discharge parameters. For the theoretical estimation of $C_{da}$, the wave dispersion relation \cite{sounddaw}\cite{raodaw1} in the case of cold dust ($T_d$ = 0) and long wavelength limit ($(K^2 \lambda_D^2 << 1)$) gives the phase velocity $C_{da}$ = $
Z_d\left({k_B T_i}{n_{d0}}/{m_d}n_{i0}\right)^{1/2}$ , where $K$, $\lambda_D$, $k_b$, $T_i$, $m_d$, 
$n_{d0}$, $ n_{i0}$, and $Z_d = Q_d/e$ are the wave vector, the dust Debye length, the Boltzmann constant, ion temperature, dust 
particle mass, equilibrium dust density, ion density and dust charge number, 
respectively\citep{sounddaw,raodaw1,dawmerlino} The theoretical estimated phase velocity $C_{da}$ of the DAW is found 
to be $\sim$ 16 $mm/s$ for the parameters $T_e \sim$ 3 eV, $T_i = 0.025$ eV, $m_d$ $\sim$ $6\times10^{-15}$ kg, $n_{d0}\sim$ 8$\times$ $10^5$ $cm^{-3}$, $n_{i0}\sim$ $9\times$ $10^8$ 
$cm^{-3}$, and $Z_d \sim$ 2 $\times$ $10^3$. Both the theoretically estimated and experimentally measured phase 
velocity of DAWs are in good agreement, which confirms the acoustics nature of the self-excited waves in the dusty plasma.\par 
\begin{figure*}[h]
 \centering
\includegraphics[scale=0.80]{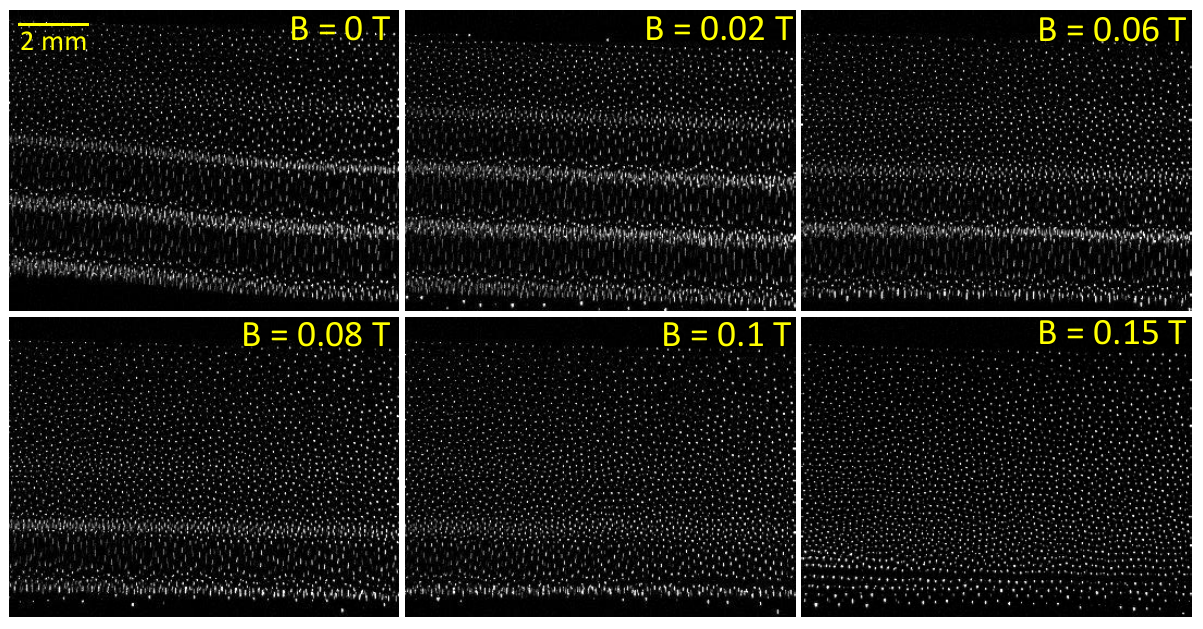}%
 \caption{\label{fig:fig3} Dust-acoustics waves in a vertical (Y--Z) plane at center (X = 0) in the presence of different strengths of external magnetic field (B). Experiments are performed at argon pressure p = 27 Pa and input rf power P = 3.5 W.} 
 \end{figure*}
\begin{figure*}[h]
 \centering
\includegraphics[scale=0.850]{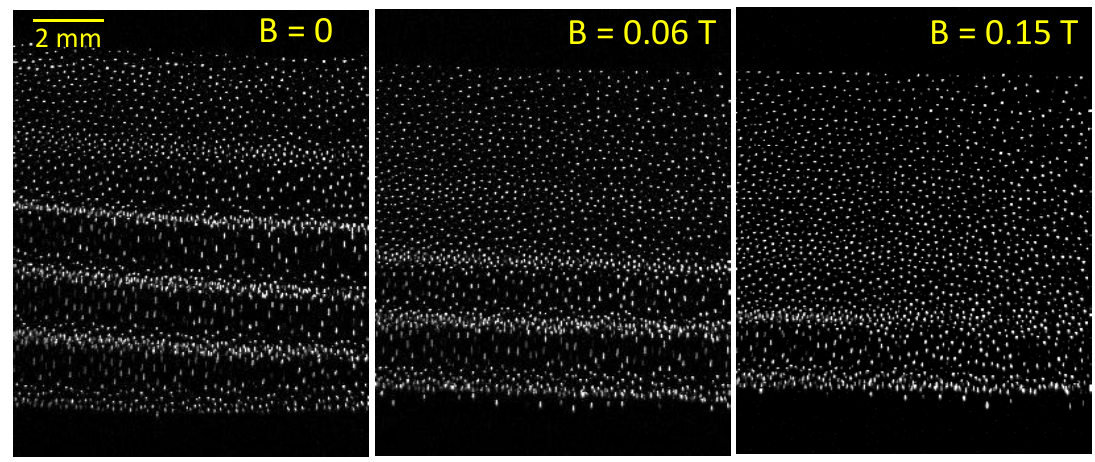}%
 \caption{\label{fig:fig4} Dust-acoustics waves in a vertical (Y--Z) plane at center (X = 0) in the presence of different strengths of external magnetic field (B). The argon pressure and rf power are set to 26 Pa and 3 W respectively for the experiements.} 
 \end{figure*}
\section{Characteristics of DAW in the presence of magnetic field} \label{sec:daw with B}
In Section~\ref{sec:DAW without B}, the excitation of DAWs and its propagation characteristics in the absence of an external magnetic field, B = 0 T have been discussed. The effect of an external magnetic field on the propagating DAWs, which are excited at p = 27 Pa and P = 3.5 W, is shown in Fig.~\ref{fig:fig3}. At this discharge conditions, the excited waves are slightly unstable at B = 0 T and get stabilized at a low magnetic field (B $<$ 0.05 T). Only the propagation characteristics such as frequency, wavlength and velocity of the DAWs get modified at low B range. Further increase of B (B $>$ 0.05 T) starts to damp the DAWs in the upper part of the confined dusty plasma medium. In the range of B, 0.05 T up to 0.13 T, the DAWs continuously damp from top to bottom of the dust grain medium. At higher B $\sim$ 0.13 T, the DAWs are found to be completely damped and the dust particles do not show any oscillatory motion around their equilibrium position. Further increase in B (B $>$ 0.13 T) turns the central region of the dusty plasma in a E$\times$B motion \cite{rotationknopka1,rotationkarasev2} which is not the focus of the present study and therefore not considered. It should be noted that dust grains show a rotational motion or E$\times$B motion due to the radial electric field. The magnitude of this E field decreases from the edge region to the central region of the lower electrode. Large diameter of the lower electrode provides a weak radial E-field at the central region of the dust cloud, therefore, the oscillatory motion of dust grains (along B) dominate over the E$\times$B motion (perpendicular to B). Hence, the effect of radial E-field during the wave motion is observed to be insignificant.\par
It has been discussed the excitation and stability of DAWs dependence on the $E/p$ ratio. We observe the characteristics of DAWs at a higher value of $E/p$, which is possible at slightly lower power (P = 3 W) and lower pressure (p = 26 Pa). Fig.~\ref{fig:fig4} represents the propagation characteristics of DAWs at various strength of the magnetic field. At this discharge conditions, the DAWs are slightly unstable compared to the previous discharge case (Fig.~\ref{fig:fig3}) in unmagnetized plasma. In the range of magnetic field B $<$ 0.06 T, the dusty plasma medium gets stabilized and the wave parameters change. With increasing B from 0.06 T to 0.15 T, we see the damping of DAWs in the upper part of dusty plasma. In this case, the DAW are not completely damped in the lower part of dusty plasma even at the higher magnetic field (B = 0.15 T). The role of B on the damping of the wave becomes negligible after a certain value. Therefore, the bottom part of the dusty plasma medium exhibits a wave like motion even after increasing the magnetic field, B $>$ 0.15 T. It is also found that higher magnetic field is required to damp the wave when the rf power is lowered. \par
\begin{figure*}[h]
 \centering
  \includegraphics[scale=0.85]{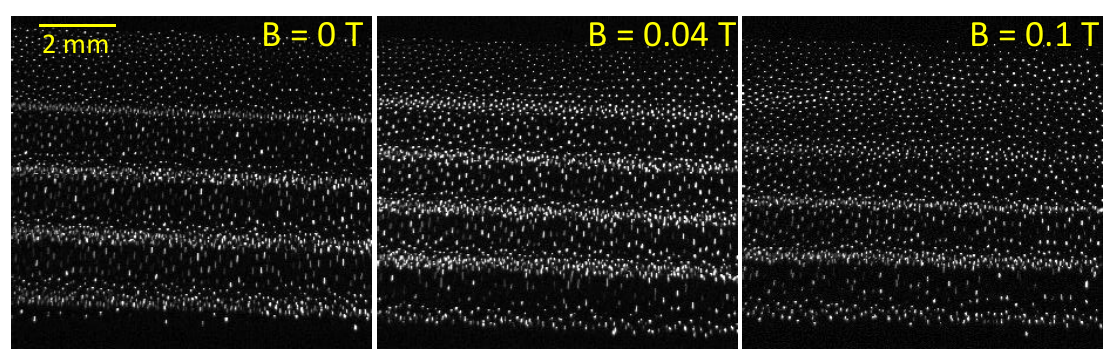}%
 \caption{\label{fig:fig5} Dust-acoustics waves in a vertical (Y--Z) plane at center (X = 0) in the presence of different strengths of external magnetic field (B). The argon pressure and rf power are set to 23 Pa and 3 W, respectively.}
 \end{figure*}

Further lowering of the pressure to 23 Pa at P = 3 W, the DAWs are observed to be unstable and stabilize in the presence of a magnetic field. Fig.~\ref{fig:fig5} represents the DAWs at different strength of B. It is found that DAWs do not damp even at the higher magnetic field (B $>$ 0.1 T) at this discharge conditions. The magnetic field only modifies the characteristics of the propagating DAWs or wave parameters. The wave parameters such as $\lambda$, $C_{da}$ and $f_d$ are measured for different B. The wave parameters with various strength of B are listed in table I. It is found that wavelength of DAW first decreases at lower B (B $<$ 0.05 T) and after that its value slightly increases again towards higher B (B $>$ 0.1 T). The opposite trend is seen in the case of the velocity and frequency of DAWs. The $C_{da}$ and $f_d$ increases with increasing the B (B $<$ 0.05 T) and after that, both values start to decrease towards higher magnetic field. Similar kind of variation in wave parameters is also observed for other discharge conditions (P = 3.5 and 3 W, p = 27 and 26 Pa). Thus, the results clearly show the modification and damping of dust-acoustic waves in the magnetized rf discharge plasma.
\begin{figure*}[h]
 \centering
\includegraphics[scale=0.50]{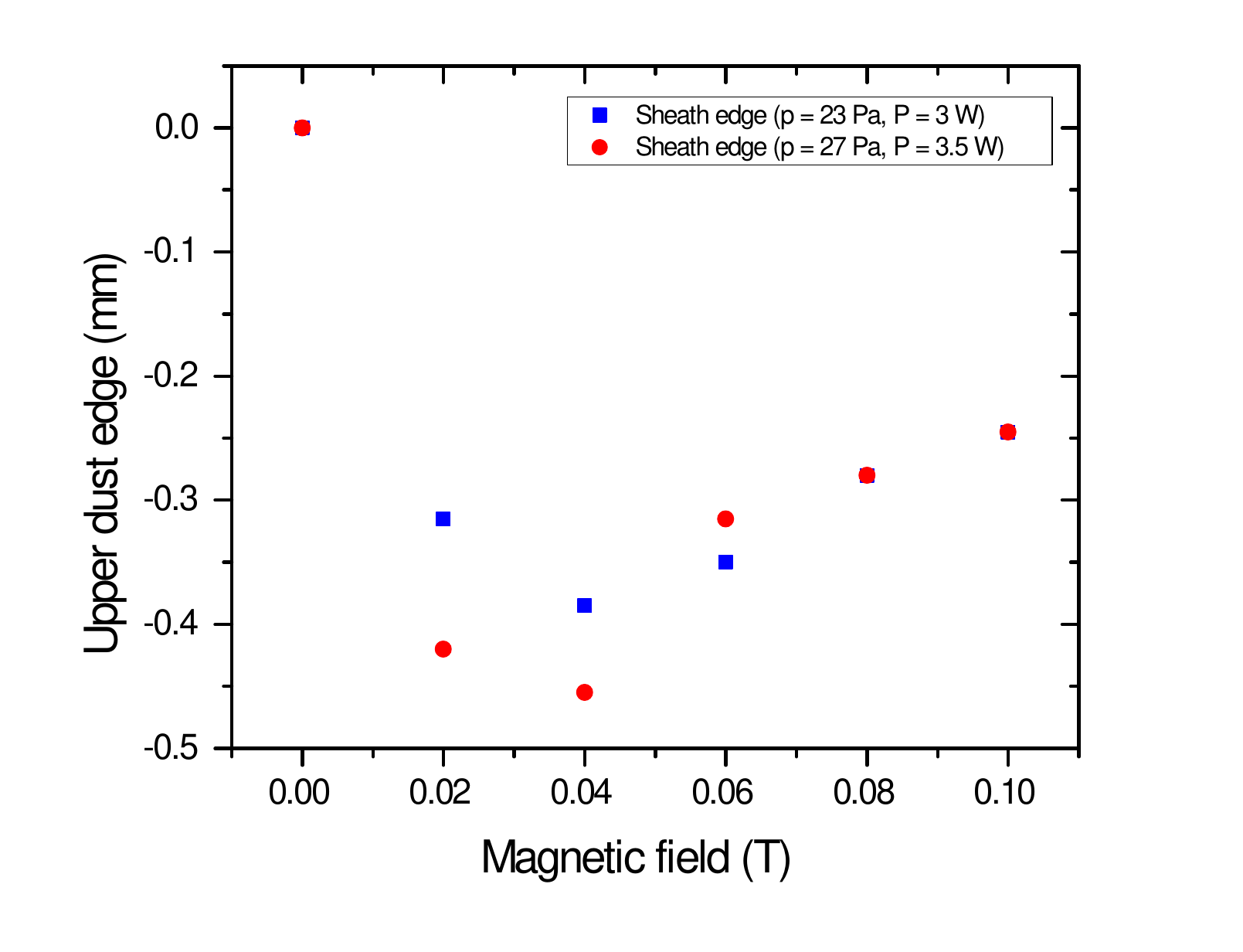}%
 \caption{\label{fig:fig6} Relative variation of upper edge particles (or sheath edge particles) with the external magnetic field for discharge conditions of Fig.~\ref{fig:fig3} and Fig.~\ref{fig:fig5}. The negative value represents a suppression of the dust cloud or shifting of upper edge to a lower position. The error in calculated values are within $\pm$ 10 \%} 
 \end{figure*}
\begin{table*}
\caption{Dust acoustic wave characteristics with magnetic field.  The experiments are performed at fixed gas pressure p = 23 Pa and rf power P = 3 W .}
\centering 
\begin{tabular}{||l|c|c| c||}
\hline
\hline 
Magnetic field &Wavelength              &Phase 
velocity             & Frequency \\
     B in T & ($\lambda$) in mm  & 
($v_{ph}$) in mm/sec  & ($f_d$ = $v_{ph}$/$\lambda$) in Hz\\
\hline
0 & 1.90 $\pm$ 0.5 & 10.2 $\pm$ 0.3  &  5.5 $\pm$ 0.4\\   
   0.02 & 1.63 $\pm$ 0.4 & 13.12 $\pm$ 0.3 &8.2$\pm$ 0.3\\
    0.04  & 1.55 $\pm$ 0.4 & 13.45 $\pm$ 0.3& 8.7 $\pm$ 0.3\\   
   0.06  & 1.59 $\pm$ 0.4  & 13.12 $\pm$ 0.5  &8.2 $\pm$ 0.4\\
     0.1  & 1.66 $\pm$ 0.4  & 12.34 $\pm$ 0.3  & 7.4$\pm$ 0.3\\
\hline \hline 
\end{tabular}
\end{table*}
\section{Discussion} \label{sec:discussion}
In the present sets of experiments, discharge is ignited between electrodes using a 13.56 MHz rf source. In such high-frequency discharge, the ion frequency ($\omega_{pi}$) is lower than the rf frequency ($\omega_{rf}$). Therefore they are not able to react to the fast changing rf electric field. It means that ions only respond to the E-field averaged over an rf cycle, resulting in a constant flux of ions to the electrodes. However, the electron frequency ($\omega_{pe}$) is higher than $\omega_{rf}$. Hence electrons follow the E-field during the rf cycle and oscillate between the electrodes on the background of the positive space charge of ions. The energetic electrons gain energy in the oscillating sheath E-field and lose energy via collisions with neutral gas resulting in a net energy loss of electrons sustaining the discharge through the ionization mechanism \cite{libermanprocessingbook,rfdischarge2}. It should be noted that during the rf period, a constant ion flux is lost from the bulk plasma to the electrodes. At each electrode, the ion current ($I_i$) is compensated by the electron current ($I_e$) that leave the plasma only at the time of the sheath collapse in a rf cycle at the respective electrode. When the plasma equilibrium is achieved, the rate of change of plasma density ($n$) which depends on the ionization rate ($R_I$) and loss rate ($R_L$) is zero, i.e, $dn/dt = R_I - R_L = 0$. The ionization rate depends on the electron density, electron energy, and ionization frequency. The plasma loss rate is mainly defined due to the ambipolar diffusion of plasma species to the chamber wall. The rate of change of electrons mainly depends on the input power, ionization rate, and the internal energy of electrons\cite{libermanprocessingbook,plasmaloss1,rfdischarge2,rfdischarge3}.\par
As dust particles are introduced in the plasma volume, the energetic electrons are lost to their surface and make the dust negatively charged \cite{dustcharginggoree}. Since electrons are first lost to the particle surface to reduce the electron flux and enhance the ion flux, the density of free electrons is less than the ion density \cite{electrondepeletation,dustydischarge2}. In other words, dust particles provide an extra surface area in addition to the chamber wall to the plasma losses through ambipolar diffusion. More dust particles into the confining potential well reduce more free electrons or enhance the plasma loss rate. At this condition, an accelerating E-field is necessary \cite{instability2} to compensate for the electron loss rate or to keep the current constant. Thus, the time-averaged E-field of sheath region where dust particles are levitated will be higher (Fig.~\ref{fig:fig2}(III)) than for the case of less particles (Fig.~\ref{fig:fig2}(I)). Since the ions are drifting through the dust cloud, the drift velocity, $v_i = \mu_i E$, also increases in the higher E-field. It has been discussed that the drifting ions relative to the stationary dust particles cause the instabilities, which excites the low-frequency acoustic modes in a dusty plasma. It has been observed that excitation of waves is possible when the ion drift velocity crosses a threshold value which is higher than the ion thermal speed \cite{instability2,instability3,instability1,instability5}, i.e., $v_i > v_{ith}$. For a fixed ion mobility, $\mu_i$, the drift velocity increases linearly with the E-field. Therefore, the DAWs are excited after adding more dust particles in a potential well of the rf sheath at given power and pressure, as is shown in Fig.~\ref{fig:fig2}(III).\par
It is a challenge to diagnose the background plasma of the dust grain medium which is necessary to understand the dynamics of a volumetric dusty plasma. It has been discussed that dust grains absorb free electrons and reduce the density of free electrons in a plasma. A dimensionless parameter, which is called Havnes parameter \cite{havnes,nanodusty2} $P_h$ = $Z_d n_d/n_i$ decides the density of free electrons in a dusty plasma. It has been assumed that plasma parameters are not strongly affected by the dust grains if $P_h < 1$ and plasma parameters can be used to estimate the dusty plasma parameters. In the case of high dust density, $P_h > 1$, density of free electrons is very low or one can consider the electron depleted dusty plasma. Recently, a new technique has been proposed to diagnose the dusty plasma using the dispersion relation of excited waves in a nanodusty plasma \cite{nanodusty2,nanodustymagnetized1}, where the havnes parameter is high ($P_h > 1$). In the present work, havnes parameter $P_h$ has the value between 0.1 to 0.3 for $n_d$ $\sim$ 3$\times 10^{10}$ to 9 $\times 10^{10}$ $cm^{-3}$, $n_i$ $\sim$ 6$\times 10^{14}$ to 8$\times 10^{14}$ $cm^{-3}$, and $Z_d \sim$ 2$\times$ $10^3$. In such dusty plasma ($P_h < 1$), density of free electrons play a dominate role to determine the dust charges, sheath E-field, sheath thickness etc. Therefore, we find difficulty to use this new proposed technique to estimate the dust charge and electric field in the presence of external magnetic field. It has been noticed in earlier studies as well in present study that the electrostatic probes modify the dynamics of dust grains in a local region \cite{mangilalpop,void,circulation} and can not be used to measure the plasma parameters and E-field during the wave motion. Since DAWs are excited due to adding more dust particles, plasma parameters without dust grains are not sufficient to explain the observed results. Therefore, observed results are qualitatively understood on the basis of dust dynamics in the magnetized rf plasma. As the magnetic field is applied, first it reduces the loss rate of the energetic electrons to the chamber wall due to the reduction of gyroradius. At this discharge conditions, electrons are magnetized even at low B (B $<$ 0.01 T) and ions remain unmagnetized below B $<$ 0.15 T. A fraction of these free electrons are lost to the dust grains and the remaining electrons gain energy in the time varying E-field of sheath and lose their energy through collisions with neutrals. Hence, the plasma density is expected to increase with increasing the magnetic field. There is an another possible mechanism to increase the free electron density  ($n_e$) or plasma density ($n$) by applying a B-field. The magnetic field can also reduce the electron flux to the dust grain due to the cross-field diffusion \cite{chendiffusioncoefficient}. Since the radius of particles is 1 $\mu$m, the electron current flowing along B to dust particles will not be affected by B \cite{tsytovich_2003}. However, it is expected to decrease the electron current perpendicular to B which reduces the electron loss to dust grains. A reduction of the electron flux to the dust surface in a strong B has been confirmed by the numerical simulation \cite{dustchgargingbieee2019}. These additional free electrons involve in the ionization process to increase the plasma density. Hence, both mechanisms continuously increase the plasma density or free electron density in the dusty plasma volume with increasing magnetic field. It is also expected that the first mechanism to increase $n_e$ dominates over the later mechanism at lower B (B $<$ 0.05 T). This increment in $n_e$ reduces the sheath thickness at lower B (B $<$ 0.05 T), resulting in a reduction of the volume occupied by the dust grains. In a 2D (Y--Z) plane, the shifting of upper edge particles to a lower position, as presented in Fig.~\ref{fig:fig6} is an indication of changing the sheath thickness. It is assumed that upper edge of dust cloud represents the sheath edge. Previous studies report a higher electron temperature ($T_e$) in the dusty plasma volume \cite{electrondepeletation, dustydischarge2}, which is expected due to the more energy gain by the free electrons in the time-varying sheath E-field. Therefore, a higher $T_e$ is also expected in our experiments which can a cause of the sheath expansion. However, the role of $T_e$ remains less effective than that of $n_e$ in the lower B range. It has been discussed that the cause of instability is due to the loss of free electrons in a dusty plasma. As the density of free electrons start to increase in the dusty plasma, the accelerating E-field starts to decrease. Since the stability of DAWs depends on the E/p ratio, which increases with lowering the ratio by reducing E-field at constant pressure (p = 27 Pa). Therefore, the stable DAWs propagate in the lower magnetic field range (B $<$ 0.05 T). Further increase in magnetic field, B $>$ 0.05 T, increases the plasma density and electron temperature via ionization process in the time-varying E-field of the sheath. The increase in $n_e$ should further suppress the dust grain medium but we observe a slight expansion of the dust cloud (see Fig.~\ref{fig:fig6}).  Shifting of the upper edge particles to a higher position indicates either the expansion of rf sheath or expansion of dust cloud. Expansion of the sheath is possible due to the dominant role of $T_e$ rather than $n_e$ in the higher B range. This variation in $T_e$ may also increase the charge $Q_d$ on the dust particles \cite{chargingbarken}. Thus, the shifting of particles to higher position is an indication of increase in $T_e$ in the range of B 0.05$\geq$ B $\leq$ 0.13T. Since the increase in $n_e$ and $T_e$ with B continuously lowers the sheath E-field, the DAWs start to be damped after B $>$ 0.05 T and are completely damped at higher B $\sim$ 0.13 T (see Fig.~\ref{fig:fig3}). Damping of DAWs is possible due to the suppression of instabilities associated with streaming ions in the sheath electric field. Since external magnetic field lowers the sheath E-field, drift velocity ($v_i = \mu_i E$) decreases with lowering the E-field and hence, DAWs get damped at higher B. There is a limited theoretical work which reports the DAWs in a magnetized dusty plasma \cite{dawmagnetized1,dawmagnetized2}. Salimullah et al. \cite{dawmagnetized1} have investigated the role of magnetic field on the DAWs in the plasma and observed the modification as well as damping of DAWs in the presence of magnetic field. Hence, the theoretical studies also support our experimental investigations.\par
In the case of low power and pressure (P = 3 W and p = 23 Pa) dusty plasma, ultimately we are reducing the ionization rate, therefore, relative higher E-field is required to keep the current constant or to balance the plasma loss rates. This higher accelerating E-field and lower ion mobility at lower pressure increase the E/p ratio and the dusty plasma becomes more unstable. After turning on the B, the reduction in electron loss rates to the wall as well as dust particles enhances the ionization rate which lowers the E-field. Therefore, the propagating characteristics of DAW as well as its stability change with increasing B. Since the ion mobility does not depend on B, its value remains unchanged in the presence of B-field. Therefore, the medium support the DAWs even at higher magnetic field (Fig.~\ref{fig:fig4} and Fig.~\ref{fig:fig5}). The modification of the wave parameters in the presence of B at P = 3 W and p = 23 Pa is shown in Table I. The phase velocity ($C_{da}$) of DAWs depends on $Q_d$, $n_d$, and $n_i$. The $n_d$, $n_i$ and $Q_d$ increases with magnetic and the combined effect of all these variables increases $C_{da}$ at lower B (B $<$ 0.05 T). Towards higher B, the dust cloud expansion decreases $n_d$ and we also expect a sight higher value of $n_i$. These both variables may decide the value of $C_{da}$. Similarly, the change in wavelength can also be understood on the basis of dust cloud suppression and expansion in the presence of B. Hence, the basic properties of an rf discharge and  dusty plasma help to understand the stability and damping of DAWs in the presence of an external magnetic field.

\section{Conclusion}           \label{sec:summary}
This paper highlights the results of  dust acoustics waves in an rf plasma and its propagation characteristics in the presence of an external magnetic field. The dusty plasma is created by introducing MF particles of 1.025 $\mu$m radius into the electrostatic potential well over the lower electrode created by the rf sheath field. The dusty plasma medium supports low-frequency acoustics waves after adding more particles in the potential well. The instability due to the streaming ions through the dust grain medium provides sufficient energy to grow such modes. These self-excited waves propagate in the direction of drifting ions and have a characteristics similar to the sound waves or acoustic waves. The characteristics of the propagating dust-acoustics waves (DAWs) get modified while the external magnetic field is introduced. For moderate pressure and power (P = 3.5 W, p = 27 Pa), the lower magnetic field (B $<$ 0.05 T) modifies the propagation characteristics of DAWs and the waves get damped at the higher magnetic field (B $\sim$ 0.13 T). On the other hand, the magnetic field only stabilizes and modifies the characteristics of propagating DAWs while the power and pressure are lowered to 3 W and 23 Pa, respectively. The stability and damping of DAWs are qualitatively explained on the basis of the change of $E/p$ ratio in the presence of a magnetic field. The magnetic field lowers the E-field of sheath where particles are confined at given discharge conditions. At weak E-field streaming ions do not cross a threshold drift velocity to excite the oscillatory motion of dust particles through the instabilities. But in lower pressure, the ion mobility is higher which increases the ion drift velocity even in a weak E-field. Therefore, the dust grain medium exhibits a wave like motion. In the dusty plasma, one of the challenges is to diagnose the background plasma in the presence of dust particles. Probe techniques are not reliable to use in a dusty plasma because probes either perturb the dusty plasma or modify its characteristics. The use of spectroscopy also becomes complicated in the case of dusty plasma with an external magnetic field. Therefore, numerical simulations of dusty plasmas in the presence of a magnetic field is required for detailed study of DAWs in the magnetized rf plasma, which will be the scope of our future work.

\section*{Acknowledgement}
This work is supported by the Deutsche Forschungsgemeinschaft (DFG). The authors are grateful to Christopher Dietz and Benjamin Steinmueller for experimental assistance.
\bibliography{aipsamp}
\end{document}